\documentclass[%
 aip,
 amsmath,amssymb,
 reprint,%
 citeautoscript,
]{revtex4-1}

\usepackage{graphicx}
\usepackage{dcolumn}
\usepackage{bm}

\usepackage[utf8]{inputenc}
\usepackage[T1]{fontenc}
\usepackage{mathptmx}
\usepackage{color}
\usepackage{hyperref}
\usepackage{amsthm}
\newtheoremstyle{query}%
{}{}
{\color{red}}
{}
{\sffamily\bfseries}{:}{12pt}
{}
\theoremstyle{query}

\begin{document}

\preprint{AIP/123-QED}

\title[Unsteady rheo-optical measurements of uniaxially extending liquid polymers]{Unsteady rheo-optical measurements of uniaxially extending liquid polymers}

\author{M. Muto}%
 \email[Authors to whom correspondence should be addressed: ]{muto.masakazu@nitech.ac.jp and tagawayo@cc.tuat.ac.jp}
\affiliation{ 
Department of Mechanical Systems Engineering, Nagoya Institute of Technology, Gokiso Campus 3-502, Gokiso, Showa-ku, Nagoya-shi, Aichi, Japan
}%
\author{Y. Tagawa}%
 \email[Authors to whom correspondence should be addressed: ]{muto.masakazu@nitech.ac.jp and tagawayo@cc.tuat.ac.jp}
\affiliation{ 
Department of Mechanical Systems Engineering, Tokyo University of Agriculture and Technology, Koganei Campus 6-204, 2-24-16 Nakacho, Koganei-city, Tokyo, Japan
}%

\date{\today}

\begin{abstract}


The viscoelastic behavior of liquid polymers changes depending on the microstructure of the polymer chains.
To investigate the unsteady structural changes induced by polymer chain morphology under stress, we develop a simple rheo-optical technique combining a high-speed polarization camera and dripping-onto-substrate capillary breakup extensional rheometry (CaBER-DoS), which uniaxially extends and aligns the polymer along the stress direction.
Using this technique, the viscoelastic behavior and birefringence field of an extending liquid polymer are simultaneously measured.
Unlike the conditions that occur under shear-stress loading with a rotational rheometer, this technique provides a linear relation between birefringence and extensional stress.
The measurements reflect the structural change in a flexible polymer (a mixed solution of PEO and CNCs) in the temporal evolution of the birefringence and orientation angle, which is a measure of the orientation state and coil-stretch transition of a polymer chain.
Remarkably, within the elasto-capillary regime, the measured birefringence remains constant with a constant orientation state, while the Weissenberg number gradually increases, which provides experimental evidence of the coil-stretch transition of the polymer chain under constant extensional stress loading.

\end{abstract}

\maketitle


\section{\label{introduction}Introduction\protect}
For a liquid polymer, the stress and velocity distributions of the fluid are complicated owing to the effects of flow resistance and apparent viscosity induced by the influence of the microstructure of polymer chains.
To investigate the phenomena occurring in liquid polymers, rheo-optical techniques\cite{noto2020applicability,philippoff1956flow,lodge1955variation,lodge1956network,shikata1994rheo,gortemaker1976re,yevlampieva1999flow,meyer1993investigation,chow1984response,ito2016shear} involving simultaneous rheometric and optical measurements are often used.
A variety of rheo-optical techniques combining a rotational rheometer, which applies deformation to the fluid with a moving wall, and birefringence measurement have been reported.
With these techniques, changes in microstructure, indicated by the orientational state and coil-stretch transition (the degree of stretch from the coil state) of polymer chains, can be inferred from birefringence measurements.
By matching results at different scales measured separately by rheometry and an optical method, a comprehensive evaluation of the phenomena occurring in liquid polymers can be performed.
While there have been many studies of the microstructural behavior of liquid polymers under ``shear'' conditions\cite{ito2016shear,haward2019flow,zhao2016flow,haward2012extensional,iwata2019local}, there have been few reports on changes in the microstructure under extensional conditions. 

To investigate polymer chain morphology under coil-stretch transition, the Weissenberg number $Wi$, which is a dimensionless number representing the ratio of elasticity and viscosity, is often used as an indicator.
However, evidence that $Wi$ is in fact an indicator of coil-stretch transition is limited to the results of numerical calculations\cite{entov1997effect,mathues2018caber}, which depend on the type of constitutive model chosen \cite{sur2018drop,mckinley2005visco,prabhakar2006effect,doyle1998relaxation,doyle1997dynamic}.
On the other hand, there have been few reports of experimental investigations of the changes in molecular structure of polymer chains under extensional conditions.
Lodge\cite{lodge1955variation,lodge1956network} and Philippoff\cite{philippoff1956flow} proposed to measure the flow-induced birefringence of highly viscoelastic solutions reeled in by a draw wheel.
Rothstein and McKinley\cite{rothstein2002comparison} developed a method for combined measurements by a filament stretching rheometer and a birefringence measurement system for dilute solutions undergoing extension.
These techniques enable investigation of the effect of extensional stress on the flow-induced birefringence of liquid polymers.
Nevertheless, to demonstrate the validity of the stress-optic law\cite{noto2020applicability,janeschitz2012polymer,ryu1996simple,inoue1991birefringence} for a flowing polymer fluid, it is important to simultaneously measure the birefringence and the orientational state of the polymer chain under extensional stress loading.
For this purpose, it is necessary to establish a simple approach to the investigation of the above-mentioned complex problems through birefringence measurements and develop a simple rheo-optical measurement method using a high-speed polarization camera.




The purpose of this study is to develop a simple system for examining changes in microstructure (i.e., the orientational state and coil-stretch transition of polymer chains) of liquid polymers under extensional stress loading with the birefringence measurements.
For this purpose, we have chosen dripping-onto-substrate capillary breakup extensional rheometry (CaBER-DoS)\cite{dinic2020flexibility, dinic2017pinch,sur2018drop,mckinley2005visco,mckinley2000extract,clasen2006dilute} as the means of applying the extensional stress.
Unlike a rotational rheometer, which measures the shear viscosity, CaBER-DoS is a noncontact method that measures the extensional viscosity from the radius of the liquid filament, which decays as the fluid is extended.
In this paper, we report the results of an investigation of the chain morphology of a polymer chain in a stretched flow field, where elasticity becomes dominant.
Selection of the elastic-dominant regime, i.e., the elasto-capillary (EC) regime is particularly important when calculating viscoelastic properties such as the extensional viscosity, strain, and relaxation time.
Importantly, within the EC regime, a uniaxial extensional stress is applied to the fluid, allowing the assumption of uniform flow without shear stress to be made\cite{mckinley2005visco}.
Therefore, the polymer chains inside the fluid are assumed to be oriented in the extensional direction and uniformly aligned, which should be experimentally demonstrable by the birefringence measurements.
Note that, according to the theory of CaBER-DoS, three regimes, dominated by inertia, elasticity, and viscosity, respectively, can be defined on the basis of the temporal decay of the radius of the liquid filament\cite{sur2018drop,mckinley2005visco,entov1997effect,wagner2015analytic,prabhakar2006effect}.
Thus, the most important requirement for our rheo-optical method is to be able to simultaneously measure the changes of the three regimes in CaBER-DoS and the birefringence, which reflects the morphology of the polymer chain.
To do this, a high-speed polarization camera, which enables us to obtain polarization information (i.e., the flow-induced birefringence and the orientation angle), is used.

\section{\label{method}Methods\protect}
In this section, the mechanism of flow-induced birefringence of a polymer and the measurement principle of the rheo-optical technique that uses both a high-speed polarization camera and the CaBER-DoS system are described.

\subsection{\label{flow-induced birefringence}Flow-induced birefringence of polymer}
The flow-induced birefringence of polymer chains is a key indicator of chain morphology.
The flow-induced birefringence $\delta_n$ that appears due to stress loading is proportional to the principal stress difference $\sigma$ in what is known as the stress-optic law \cite{noto2020applicability,janeschitz2012polymer,ryu1996simple,inoue1991birefringence}:
\begin{equation}
\delta_n = |n_{\parallel}-n_{\perp}| = C |\sigma_{\parallel}-\sigma_{\perp}|,
\label{eq2},
\end{equation}
where $n_{\parallel}$ and $n_{\perp}$ are the refractive indices for polarizations respectively parallel and perpendicular to the orientation of the polymer chain and $C$ is the stress-optic coefficient.
In our experimental setup using birefringence measurements, the intensity of the flow-induced birefringence can be measured from the polarization of the light transmitted through the liquid polymer.

\begin{figure}[!t]
\includegraphics[width=\columnwidth]{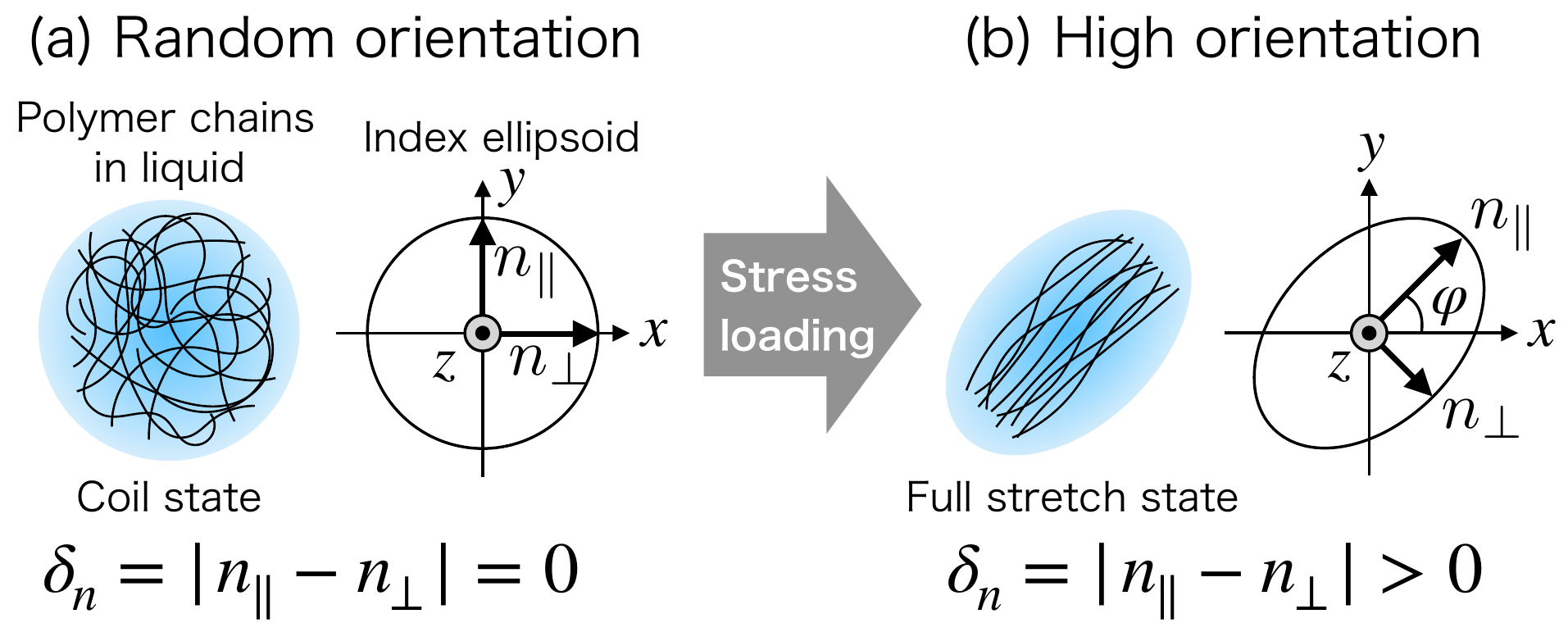}
\caption{\label{fig2} 
Flow-induced birefringence induced by a change in the orientational state of polymer chains under stress loading.}
\end{figure}

When there is no stress loading, polymer chains in the liquid have random orientations in the coil state, and thus flow-induced birefringence does not occur [see Fig.~\ref{fig2}(a)].
When stress is applied to the liquid polymer, the polymer chains become highly aligned by stretching, resulting in increased birefringence $\delta_n$ [see Fig.~\ref{fig2}(b)].
In this paper, the orientation angle $\varphi$ is defined as the tilt angle of the polymer chain with respect to the horizontal ($x$) direction [the vertical ($y$) direction is the extensional direction].

\subsection{\label{rheo-optical technique}Rheo-optical technique with high-speed polarization camera}
\begin{figure*}[t!]
\includegraphics[width=0.74\textwidth]{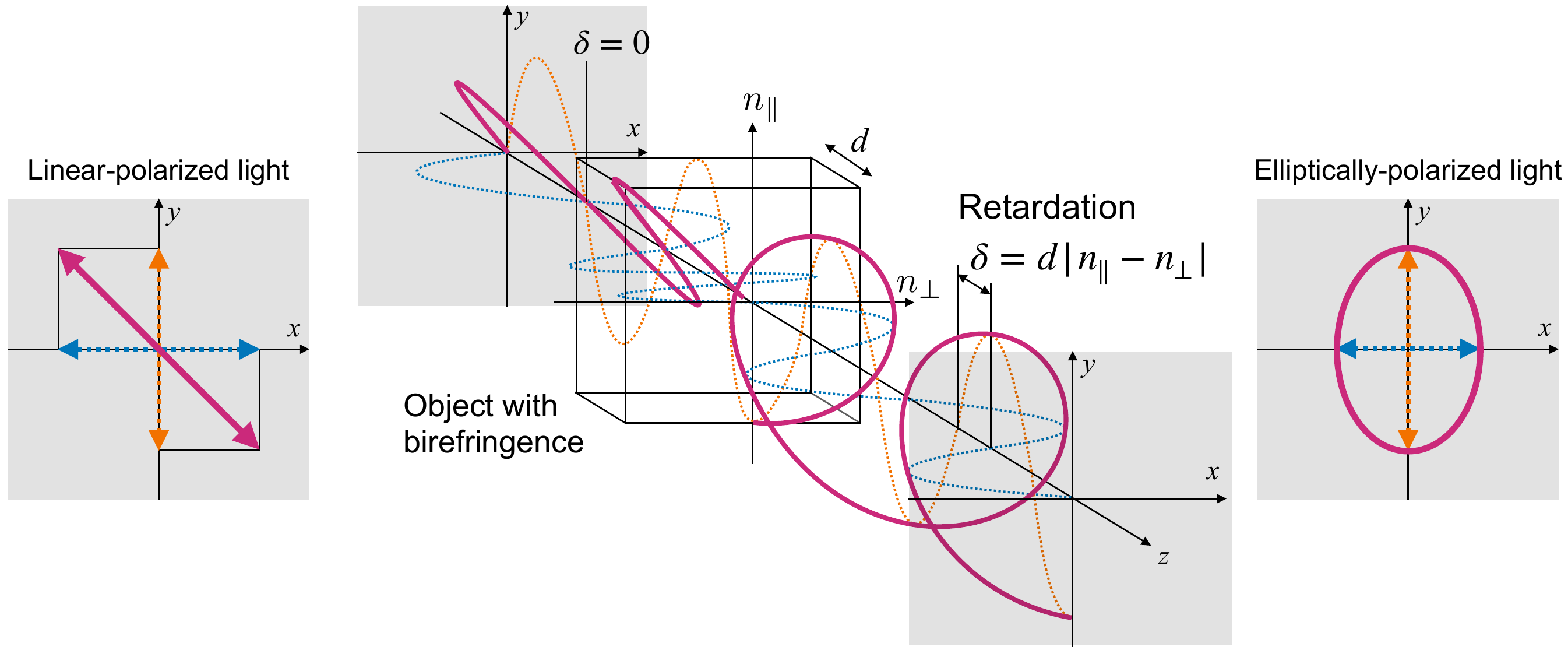}
\caption{\label{rheo_optical_technique}
Rheo-optical technique with polarization used to measure the phase retardation induced by changes in the polarization state of a birefringent object.}
\end{figure*}

We employed a birefringence measurements\cite{bass2010handbook,rastogi2015digital} to determine the phase retardation of polarized light passing thorough polymer chains under stress loading.
Figure~\ref{rheo_optical_technique} shows how the polarization state of light changes when it is transmitted through a birefringent object.

The incident electromagnetic wave, which is represented by the red curve, consists of two orthogonal components, which are represented by the blue and orange curves.
When both components are in phase, the retardation is zero, and thus the incident light is linearly polarized. 
When light is transmitted through a birefringent object (the liquid polymer) under stress loading, the propagation speed of each component wave differs, since the refractive indices $n_{\parallel}$ and $n_{\perp}$ are different from one another.
As a result, the phases of the component differ, and the values of the retardation change.
The retardation $\delta$ is given by the integral of the birefringence $\delta_n$ over infinitesimal volume elements within the object along the optical axis (the $z$ direction):
\begin{equation}
\delta = \int \delta_n\,dz
\label{eq4}.
\end{equation}

\begin{figure*}[t!]
\includegraphics[width=0.72\textwidth]{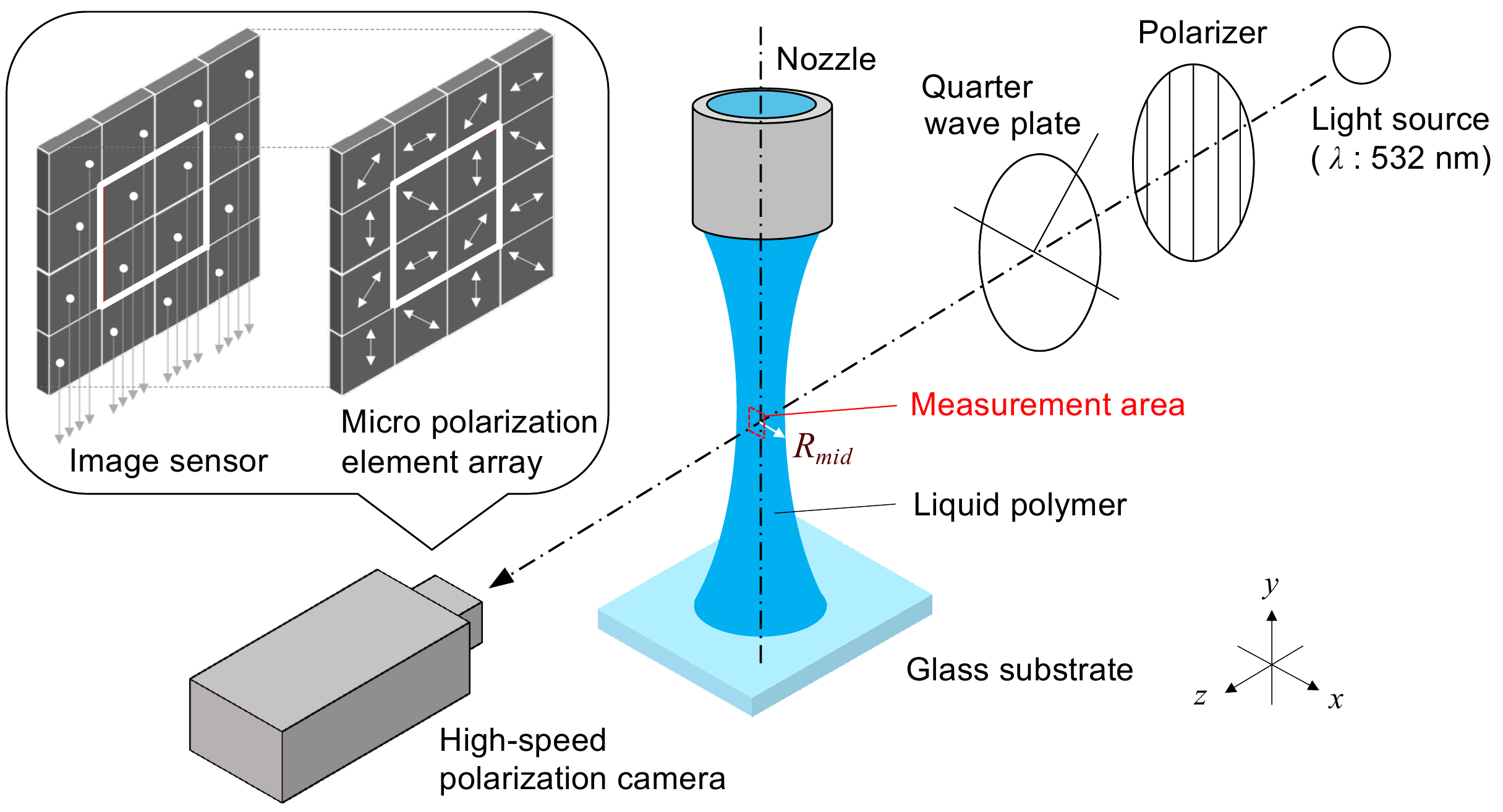}
\caption{\label{fig4}
Experimental setup using a high-speed polarization camera and CaBER-DoS system.}
\end{figure*}

We propose a simple calibration technique that combines a CaBER-DoS system and a high-speed polarization camera based on the theory of birefringence measurement (CRYSTA PI-1P, Photron Ltd.) (see Fig.~\ref{fig4}). 
Our concept can be realized by using just a simple setup: a target sample (liquid polymer) between the polarization camera and an LED light source.

In the CaBER-DoS system, a fluid dispensing device with a syringe pump (Pump 11 Elite, Harvard Apparatus Ltd.) is used to deliver a drop of the polymer solution at a relatively low flow rate onto a glass substrate placed at a fixed distance below the nozzle exit.
As the droplet slowly drips from the nozzle, a filament is formed.
By measuring the minimum necking radius $R_\mathrm{mid}$ of the filament, which becomes thinner as it extends, we can calculate the extensional viscosity, strain, relaxation time, etc.
When utilizing the CaBER-DoS system, we can assume that the orientation of all chain polymers is uniform, because the stress exerted along the optical axis is uniform.
Therefore, the retardation $\delta$ over the optical depth of the measurement object $d$ in Eq.~(\ref{eq4}) can be rewritten as follows:
\begin{equation}
\delta = d\delta_n = d |n_{\parallel}-n_{\perp}|
\label{eq5}.
\end{equation}
This equation is valid for our technique, although it is generally invalid for experiments under shear stress loading.
Note that it is valid only when the incident polarized light passes through the center of the filament, in which case we can rewrite $d = 2R_\mathrm{mid}$ in our setup.

The high-speed polarization camera is capable of capturing photoelastic phenomena with a pixelated polarizer array at a frame rate of up to $1.5 \times 10^6$~frames/s.
The array consists of four adjacent polarizers ($2\times2$) arranged in four different orientations.
The instantaneous retardation ${\delta}$ and the instantaneous orientation angle $\varphi$ can be calculated from the light intensities captured by these four polarizers ($I_1$ at 0$^\circ$, $I_2$ at 45$^\circ$, $I_3$ at 90$^\circ$, and $I_4$ at 135$^\circ$)\cite{onuma2014development}:
\begin{align}
{\delta} &= \frac{\lambda}{2 \pi} \sin ^{ -1 }{ \frac { \sqrt { { ( { I }_{ 3 }-{ I }_{ 1 } ) }^{ 2 }+{ ( { I }_{ 2 }{-I }_{ 4 } ) }^{ 2 } } }{ ({ I }_{ 1 }+{ I }_{ 2 }+{ I }_{ 3 }+{ I }_{ 4 } )/2} }
\label{delta},
\\[6pt]
\varphi &=\frac{1}{2}\tan ^{ -1 }{ \frac { { I }_{ 3 }-{ I }_{ 1 } }{{ I }_{ 2 }-{ I }_{ 4 }}}
\label{phi},
\end{align}
where $\lambda$ is the wavelength of the incident light.
In this experiment, the incident light was emitted from a green LED with a wavelength of 520~nm and was polarized by a circular polarization sheet consisting of a polarizer and a quarter-wave plate.
Note that the incident light must be directed straight into the camera.

\begin{figure}[t!]
\includegraphics[width=\columnwidth]{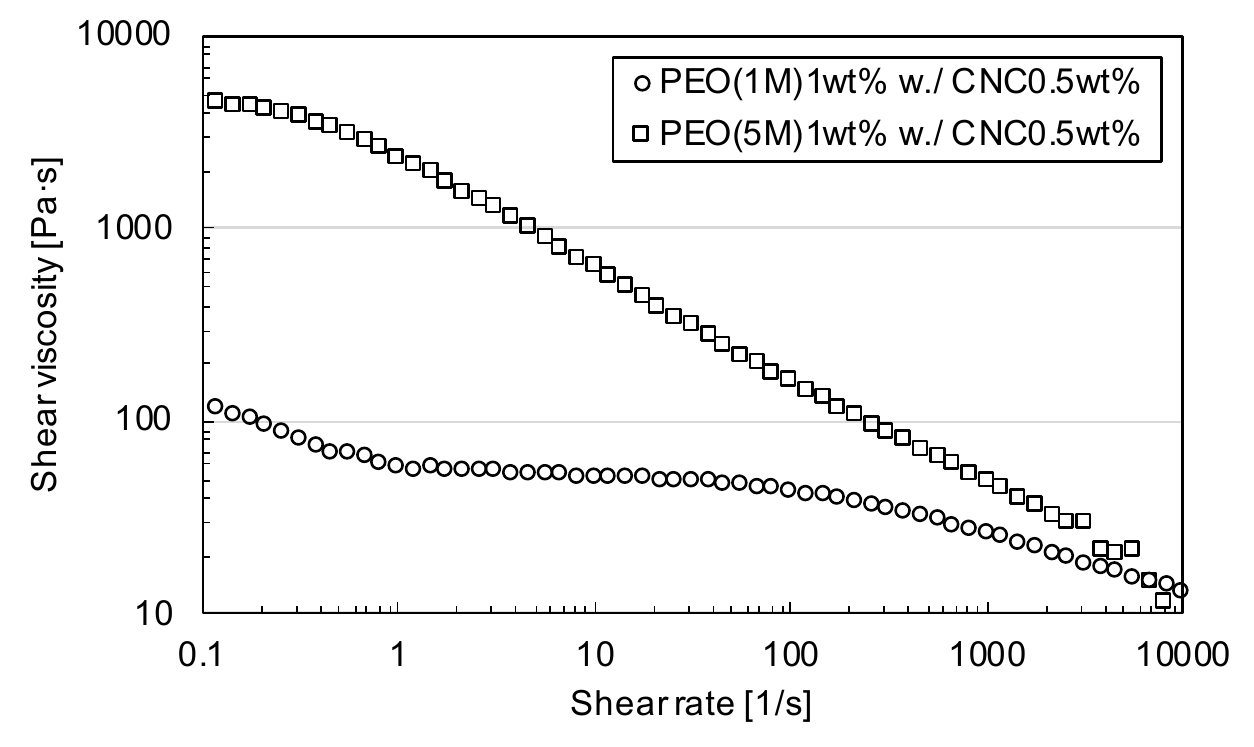}
\caption{\label{shear_PEO}
Shear-thinning viscosity curve of mixed solution of PEO and CNCs.}
\end{figure}

In this study, a mixed solution of polyethylene oxide [PEO, 1~wt.\,\% concentration in ultrapure water, molecular weight $M_{w} = (1\text{--}5)\times10^6$, Sigma-Aldrich Co. LLC] and cellulose nanocrystals (CNCs, 0.5~wt.\,\%, Cellulose Lab Inc.) was used as the liquid polymer.
The solvents were stirred in ultrapure water for 24~h with a hot stirrer, with the preset temperature and speed of revolution set to 80\,$^{\circ}$C and 1000~rpm, respectively.
Note that the mixed solutions was a non-Newtonian fluid exhibiting shear-thinning behavior (see Fig.~\ref{shear_PEO}). 

\section{\label{result}Results and discussion\protect}
We divide this section into three subsections.
The first reports the visualized phase retardation field of the extending liquid polymer filament, captured by the high-speed polarization camera.
The second reports the temporal evolution of phase retardation and orientation angle.
The third presents the flow-induced birefringence.

\subsection{\label{visualization}Visualized phase retardation field of extending liquid polymer filament}
\begin{figure*}[t!]
\includegraphics[width=0.6\textwidth]{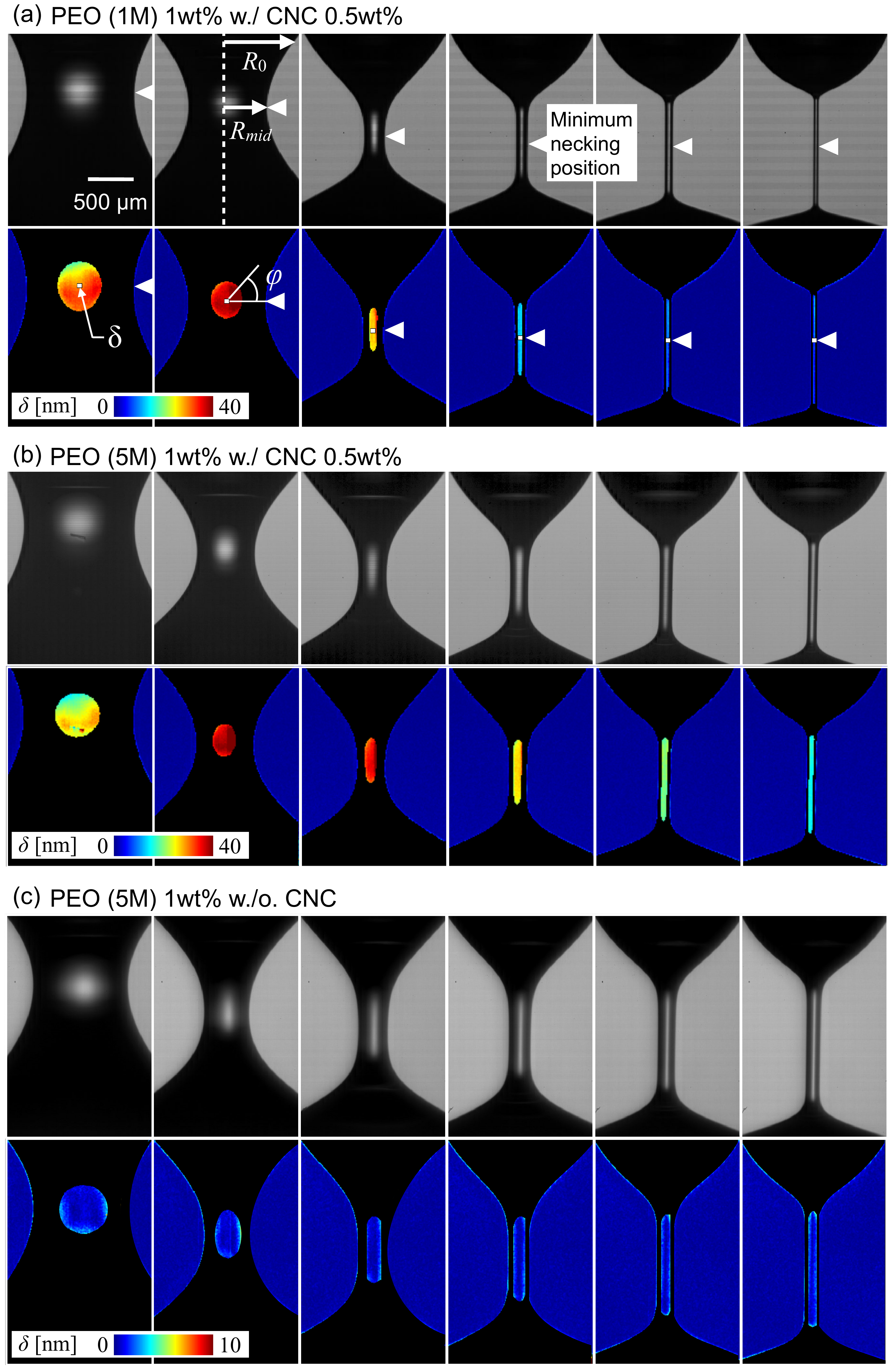}
\caption{\label{retardation field} 
Temporal evolution of the extending liquid polymer filament shown by raw images and the visualized phase retardation field for (a) and (b) mixed solutions of CNCs (0.5~wt.\,\%) and PEO (1~wt.\,\%) with molecular weights $M_{w} = 1\times10^6$ and $5\times10^6$, respectively, and (c) PEO solution ($M_{w} = 5\times10^6$, 1~wt.\,\%) without CNCs.
The white triangles indicate the positions of the minimum necking radius, and the white area ($10 \times 10$ pixels indicated by $\delta$) shows the measurement area for the retardation and orientation angle.}
\end{figure*}

The visualization results for the phase retardation field of the extending liquid polymer filament captured with the high-speed polarization camera are discussed here.
Figure~\ref{retardation field} shows the temporal evolution of the filament, using the raw images captured by the high-speed polarization camera.
The position of the minimum necking radius was also changing continually during the measurement, and could be tracked using our field measurement technique.
The retardation field was visualized by converting the intensity value at each pixel position in the raw images to the corresponding retardation value.

It was confirmed that the retardation appeared differently as the molecular weight of the PEO was changed [Figs.~\ref{retardation field}(a) and~\ref{retardation field}(b)].
This was due to the difference in the intensity of stress loading, which depends on the molecular structure and viscoelasticity of a polymer.
Moreover, in the case of the PEO solution without CNCs [Fig.~\ref{retardation field}(c)], no retardation could be observed.
Thus, we argue that the retardation in mixed solutions of PEO and CNCs is induced by microstructural changes in the CNCs, 
since the induction of birefringence by CNCs is due to their chain structure\cite{kadar2021cellulose,calabrese2021effects,hasegawa2020cellulose,arai2018transport,chowdhury2017improved,cranston2008birefringence}.

From the retardation field, as the stretching continued, the intensity of phase retardation also changed, i.e., it increased in the early stages and decreased thereafter.
This is because phase retardation is proportional not only to the principal stress difference but also to the optical thickness, i.e., the diameter of the liquid polymer filament, as indicated in Eq.~(\ref{eq5}).
For an extending liquid filament in a CaBER-DoS system, while the principal stress differences tend to increase, the diameter of the filament decreases.
Therefore, it can be predicted that retardation will decrease owing to the reduction in optical thickness.

\subsection{\label{retardation result}Phase retardation and orientation angle}
\begin{figure*}[t!]
\includegraphics[width=0.6\textwidth]{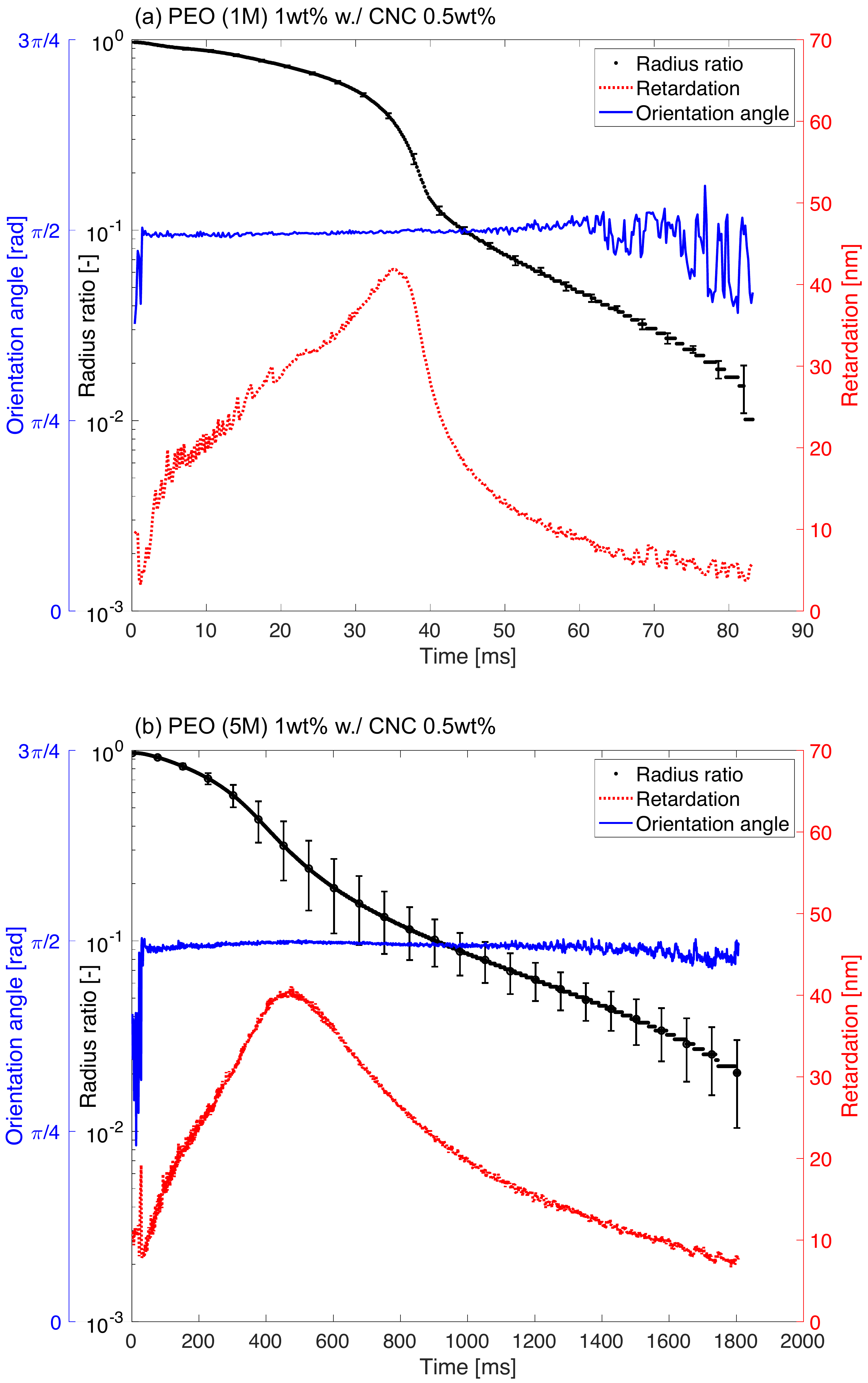}
\caption{\label{PEO} 
Temporal evolution of phase retardation $\delta$ and orientation angle $\varphi$ with radius ratio $R_\mathrm{mid}/R_{0}$ for mixed solutions of CNCs and PEO with molecular weights $M_{w} = 1\times10^6$ (a) and $5\times10^6$ (b). The error bars on the curves of the radius ratio show the standard deviation of the five measurement results.}
\end{figure*}

We now consider the measurement results for the phase retardation and orientation angle of the extending liquid polymer filament.
To investigate these quantitatively, we introduced a measurement area ($10\times10$ pixels, shown by the white box in Fig.~\ref{retardation field}) on the axis of symmetry of the filament. 
This measurement area migrated together with the minimum necking radius.
The values of the retardation and orientation angle were obtained as spatial averages over the measurement area.
Figure~\ref{PEO} shows the results for radius ratio, retardation, and orientation angle over time for two different molecular weights of PEO.

The measurement results for the temporal evolution of the radius ratio (the black plots in Fig.~\ref{PEO}) can be seen to decrease with time. 
The elapsed time for the filament to pinch off increases as the molecular weight of PEO increases.
In the elasto-capillary (EC) regime, which is an important indicator when determining the viscoelastic properties, the radius $R_\mathrm{mid}$ decays exponentially with time $t$ \cite{sur2018drop,mckinley2005visco,entov1997effect,wagner2015analytic,prabhakar2006effect}:
\begin{equation}
\frac{R_\mathrm{mid}(t)}{R_{0}} = \left(\frac{GR_{0}}{2\gamma}\right)^{1/3}\exp\!\left(-\frac{t}{3\lambda_{E}}\right)
\label{radius in ECregime},
\end{equation}
where $R_{0}$ is the initial radius, and $G$, $\gamma$, and $\lambda_{E}$ are the elastic modulus, surface tension, and relaxation time of the fluid, respectively.
Moreover, the duration of the EC regime is the same as the duration of the period when the strain rate $\dot{\epsilon}_\mathrm{mid}$ is constant: 
\begin{equation}
\dot{\epsilon}_\mathrm{mid} = -\frac{2}{R_\mathrm{mid}(t)}\frac{dR_\mathrm{mid}}{dt}
\label{strain rate}.
\end{equation}
Knowing this, we can determine the EC regime of the mixed solutions of PEO and CNCs.
Plots of the radius ratio for common viscoelastic fluids have an inflection point corresponding to the transition from the inertio-capillary (IC) regime to the EC regime \cite{sur2018drop,mckinley2005visco,entov1997effect,wagner2015analytic,prabhakar2006effect}.
The results for the mixed solutions of PEO and CNCs (Fig.~\ref{PEO}) clearly show the characteristics of the EC regime.
The plots of the radius ratio for the mixed solutions both exhibit an inflection point, and the temporal images [Fig.~\ref{retardation field}(b)] clearly show that the filament is cylindrical.
The duration of the EC regime for the mixed solutions of PEO and CNCs of different molecular weights are 46--78~ms for $M_{w} = 1\times10^6$ and 750--1730 ms for $M_{w} = 5\times10^6$.

The blue curves in Fig.~\ref{PEO} show the results for the orientation angle of the polymer chains.
Remarkably, these results confirm that the orientation angle $\varphi$ remains constant at $\pi/2$, which means that the polymer chains have become highly aligned along the extensional direction of the filament, except during the early stage (part of the IC regime) and the final stage [part of the visco-capillary regime (VC) regime].
Note that the values of the orientation angle oscillate in the VC regime owing to the lack of spatial resolution.
The relationship between the flow-induced birefringence $\delta_{n}$ and the orientation angle $\varphi$ is expressed by the following equation\cite{okada2016reliability}:
\begin{equation}
\delta_{n} = f \delta_{n}^{0} = \frac{3\cos^{2}(\pi/2-\varphi)-1}{2} \delta_{n}^{0}
\label{orientation_angle},
\end{equation}
where $\delta_{n}^{0}$ is the intrinsic birefringence, which is a material constant for the orientational birefringence of polymers.
Importantly, within the EC regime, the birefringence can be treated as intrinsic birefringence, because the orientation angle remains constant at $\pi/2$.
In such situations, the liquid polymer filament is assumed to be an axially uniform cylindrical column of constant radius $R_\mathrm{mid}$, with a free-slip boundary condition, which does not induce a radial shear flow \cite{mckinley2005visco, mckinley2000extract}.
Thus, the principal stress difference in Eq.~(\ref{eq2}) is equal to the normal stress, i.e., the uniaxially extensional stress $\sigma_{e}$, and the stress-optic law within the EC regime can be written as follows:
\begin{equation}
\delta = d \delta_n = C d \sigma_{e}
\label{eq9}.
\end{equation}
As a result, in our experiment, polymers were oriented along the extensional direction.
Thus, it is confirmed that the CaBER-DoS system provides sufficient extensional stress to orient the polymer chains.



Each of the red plots of retardation in Fig.~\ref{PEO} exhibits a maximum point, which is at the same locations as the inflection point in the corresponding plot of the radius ratio.
Thus, the positions of these maxima in the retardation plots indicate the locations where the IC regime switches to the EC regime, 
since the results for the retardation are affected by the optical thickness [see Eq.~(\ref{eq9})].
In the next subsection, we describe a more detailed validation oft the change in birefringence that is not affected by the decay.

\begin{figure*}[t!]
\includegraphics[width=0.6\textwidth]{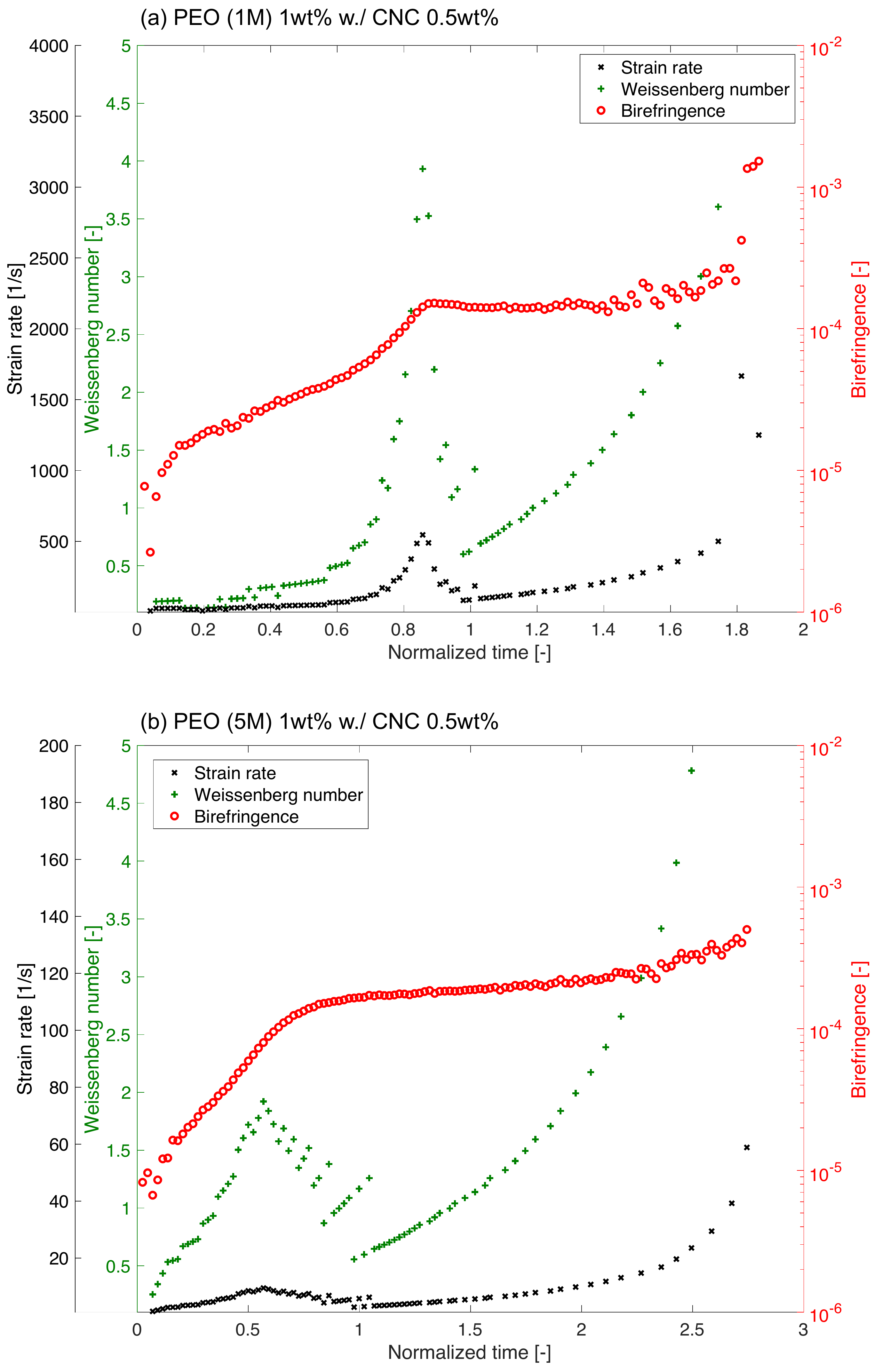}
\caption{\label{birefringence PEO} 
Temporal evolution of birefringence $\delta_{n}$, Weissenberg number $Wi$, and strain rate $\dot{\epsilon}_\mathrm{mid}$ for mixed solutions of CNCs and PEO with molecular weights $M_{w} = 1\times10^6$ (a) and $5\times10^6$ (a).}
\end{figure*}
\subsection{\label{birefringence}Flow-induced birefringence within coil-stretch transition}
We now analyze the results for the flow-induced birefringence measured with the high-speed polarization camera for an extending liquid polymer.
When all the polymer chains are aligned with the optical axis, the flow-induced birefringence $\delta_{n}$ due to the polymer chain morphology can be derived from the measured retardation $\delta$ [Eq.~(\ref{eq5})].
Figure~\ref{birefringence PEO} show plots of the birefringence $\delta_{n}$ vs normalized time $\hat{t}$ (normalized by the transition time from the IC to the EC regime).
Here, the normalized times within the EC regime for the PEO with molecular weights $M_{w} = 1\times10^6$ and $5\times10^6$ are from 1 to 1.7 and from 1 to 2.3, respectively.
These results confirm that the birefringence increases exponentially within the IC regime, has a constant value within the EC regime, and increases again within the VC regime.
Moreover, the plot of birefringence has an inflection point at the transition from the IC to the EC regime.
Within the EC regime, the variations of the birefringence with respect to the overall range are $\pm1.45\%$ and $\pm7.98\%$ for PEO with molecular weights $M_{w} = 1\times10^6$ and $5\times10^6$, respectively.
The fact that the value of birefringence remains constant within the EC regime is consistent with the theoretical basis of CaBER-DoS, according to which the uniaxial extensional stress is constant within the EC regime \cite{sur2018drop,mckinley2005visco,entov1997effect,wagner2015analytic,prabhakar2006effect}.

We believe that there are two reasons why the temporal evolution of the birefringence shows a similar trend to that of the extensional stress.
The first is that, within the EC regime, the birefringence is proportional to the extensional stress.
This is evident from the fact that Eq.~(\ref{eq9}) holds under conditions where the birefringence can be measured as the intrinsic birefringence and only a uniaxial extensional stress is exerted on the filament within the EC regime [as follows from the derivation of Eq.~(\ref{eq9}) in Sec.~\ref{retardation result}].
The second reason is that, from the IC to the EC regime, the tendency of the birefringence matches that of the capillary pressure (i.e., the extensional stress) calculated through numerical simulation by Mathues \emph{et al.}\cite{mathues2018caber}, who argued that the transition from the IC to the EC regime can be explained by the temporal evolution of the stress contributions.
Within the IC regime, the polymer chains remain in a coiled state, and the inertial acceleration balances the capillary pressure in the fluid column, which results in the strong stretching of the chains in the coiled state.
Immediately after this, the elastic stress rises quickly to balance the capillary pressure, and the inertial contribution drops to a negligible value.
The inflection point of the capillary pressure occurs when the contributions of inertial and elastic forces are swapped at the transition from the IC to the EC regime.
As a result, upon entry into the EC regime, the growth of the uniaxial deformation of the chains suddenly slows down, which results in the formation of an equilibrium state of the capillary pressure.


To verify whether the birefringence corresponds to the polymer chain morphology, we introduce the Weissenberg number $Wi$ as an indicator.
It is known that for a polymer coil to stretch, $Wi$ must at least exceed a critical value of $\frac{1}{2}$ \cite{sur2018drop,mckinley2005visco,prabhakar2006effect}.
In this study, we calculate $Wi$ as the product of the strain rate $\dot{\epsilon}_\mathrm{mid}$ and the relaxation time $\lambda_{E}$. The latter is constant \cite{mckinley2005visco}, and so 
 the curve of $Wi$ has the same shape as that of the strain rate.
Figure~\ref{birefringence PEO} show that $Wi$ remains larger than the critical value of $\frac{1}{2}$ until the pinching off of the liquid polymer filament, while the strain rate is constant within the EC regime.
Moreover, it is clear that each plot of $Wi$ has two turning points: a maximum followed by a minimum.
The normalized times at the first turning point are 0.86 and 0.57 for the PEOs with molecular weight $M_{w} = 1\times10^6$ and $5\times10^6$, respectively, both of which are within the IC region.
For both PEOs, the second turning point of $Wi$, which is at a value greater than $\frac{1}{2}$, is located at the transition from the IC to the EC regime.
The existence of two turning points, which cause the plot of $Wi$ to have a mountainous shape from the IC to the EC regime, has also been reported for other polymer solutions \cite{sur2018drop,prabhakar2006effect}.
From a comparison of the experimental results for the birefringence and $Wi$, we can confirm that the birefringence becomes constant at the point of transition from the IC to the EC regime, where the $Wi$ data show that the polymer chain gradually becomes stretched from the coiled state.
Therefore, we argue that the measured birefringence can be used as an indicator to investigate the coil-stretch transition in our rheo-optical technique that uses both a high-speed polarization camera and a CaBER-DoS system.

\section{\label{conclusion}Conclusion\protect}
To investigate unsteady changes in the microstructure of liquid polymers under extensional stress loading, we have developed a simple rheo-optical system based on CaBER-DoS and birefringence measurements.
The CaBER-DoS system was used to uniaxially extend the liquid polymer under extensional stress loading, and we successfully measured the phase retardation field of a liquid polymer (a mixed solution of PEO and CNCs) using a high-speed polarization camera.

As the molecular weight of the polymer was changed, the visualized retardation changed in appearance, owing to differences in the intensity of stress loading. 
By performing measurements in a defined area of the retardation field, we confirmed that the plot of retardation vs time exhibits a maximum point, which is at the same location as an inflection point in the plot of the radius ratio corresponding to the transition from the IC to the EC regime.
From this, we argue that the position of the maximum point of the retardation indicates where the IC regime switches to the EC regime.

According to the theoretical basis of CaBER-DoS, the direction of the extensional stress and the orientation of the polymer chain are the same within the EC regime.
To verify this, a simultaneous measurement of the orientation angle of the polymer chains was carried out, and
it was confirmed that the orientation within the EC regime is the same as the extensional direction. 
Thus, we have proved that our system enables measurement of flow-induced birefringence of polymer solutions in the case where the polymer is oriented in the extensional direction.

The plot of the temporal evolution of the flow-induced birefringence of the extending liquid polymer also exhibits an inflection point at the transition from the IC to the EC regime.
We confirmed that there is an exponential increase within the IC regime and a constant value within the EC regime; the variations in the birefringence within the EC regime with respect to the overall range are $\pm 1.45\%$ and $\pm 7.98\%$ for PEOs with molecular weights $M_{w} = 1\times10^6$ and $5\times10^6$, respectively.
This trend matches that of the uniaxially extensional stress obtained from numerical simulations in the literature and is consistent with the theoretical behavior according to which the extensional stress is constant within the EC regime.
Moreover, at the same point of transition from the IC to the EC regime, the $Wi$ data show that the polymer chain gradually becomes stretched from its coiled state.
Thus, we argue that the measured birefringence can be used as an indicator in investigations of the coil-stretch transition.

\begin{acknowledgments}
This work was supported by the Japan Society for the Promotion of Science, KAKENHI Grant Nos. 17H01246, 20H00223, 19K23483, and 20K14646.
We thank D. Yamada and Y. Matsumoto for support with the experiments.
\end{acknowledgments}

\section*{Author declarations}

\subsection*{Conflict of Interest}

The authors have no conflicts to disclose.

\section*{Data availability}

The data that support the findings of this study are available from the corresponding authors upon reasonable request.

\appendix*

\section{\label{XG} Xanthan gum solution}
\begin{figure*}[t!]
\includegraphics[width=0.6\textwidth]{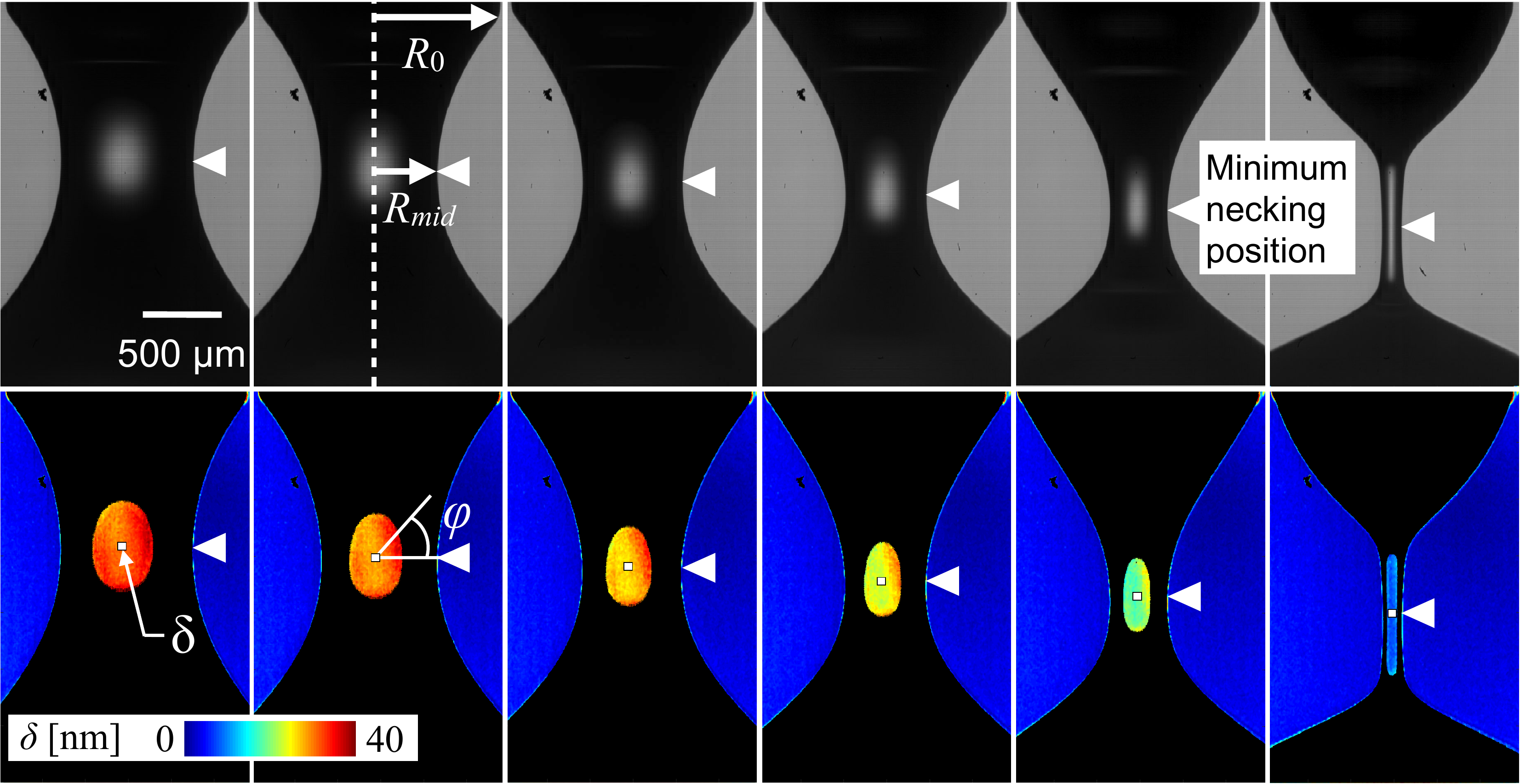}
\caption{\label{retardation field XG} 
Temporal evolution of an extending liquid polymer filament shown by raw images and the visualized phase retardation field of an XG solution with a concentration of 0.5~wt.\,\%.
The white triangles indicates the position of the minimum necking radius, and the white area ($10 \times 10$ pixels indicated by $\delta$) shows the measurement area for the retardation and orientation angle.}
\end{figure*}

\begin{figure}[t!]
\includegraphics[width=\columnwidth]{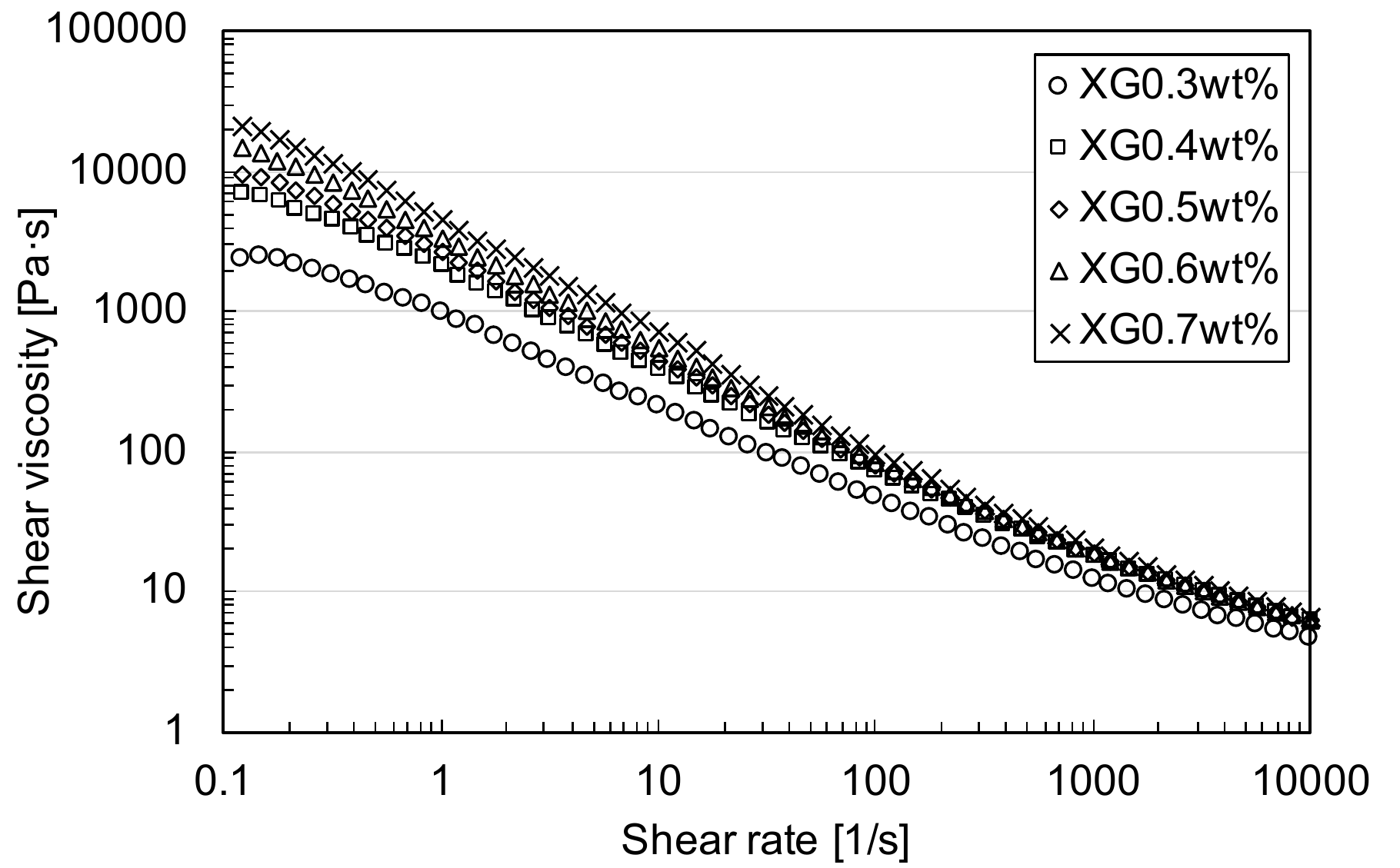}
\caption{\label{shear_XG}
Shear-thinning viscosity curve of XG solutions.}
\end{figure}

We believe that the difficulties in orienting polymer chains of polyethylene oxide (PEO) arise because PEO is more flexible than a semirigid polymer such as xanthan gum (XG) \cite{pereira2013drag}.
Thus, the fact that the polymer chains of PEO were oriented by our measurement technique could probe its generalizability.
To investigate the ease of orientation of a semirigid polymer, the same experiment as for PEO was carried out for XG solutions with various concentrations.
In this experiment (see Fig.~\ref{retardation field XG}), solutions of XG (0.3--0.7~wt.\,\% in ultrapure water, molecular weight $M_{w} = 2\times10^6$, Sigma-Aldrich Co. LLC) were used. It should be noted that XG solutions are non-Newtonian fluids that show shear-thinning behavior (see Fig.~\ref{shear_XG}) and induce birefringence due to their chain structure\cite{yevlampieva1999flow,meyer1993investigation,chow1984response}.

\begin{figure*}[t!]
\includegraphics[width=0.78\textwidth]{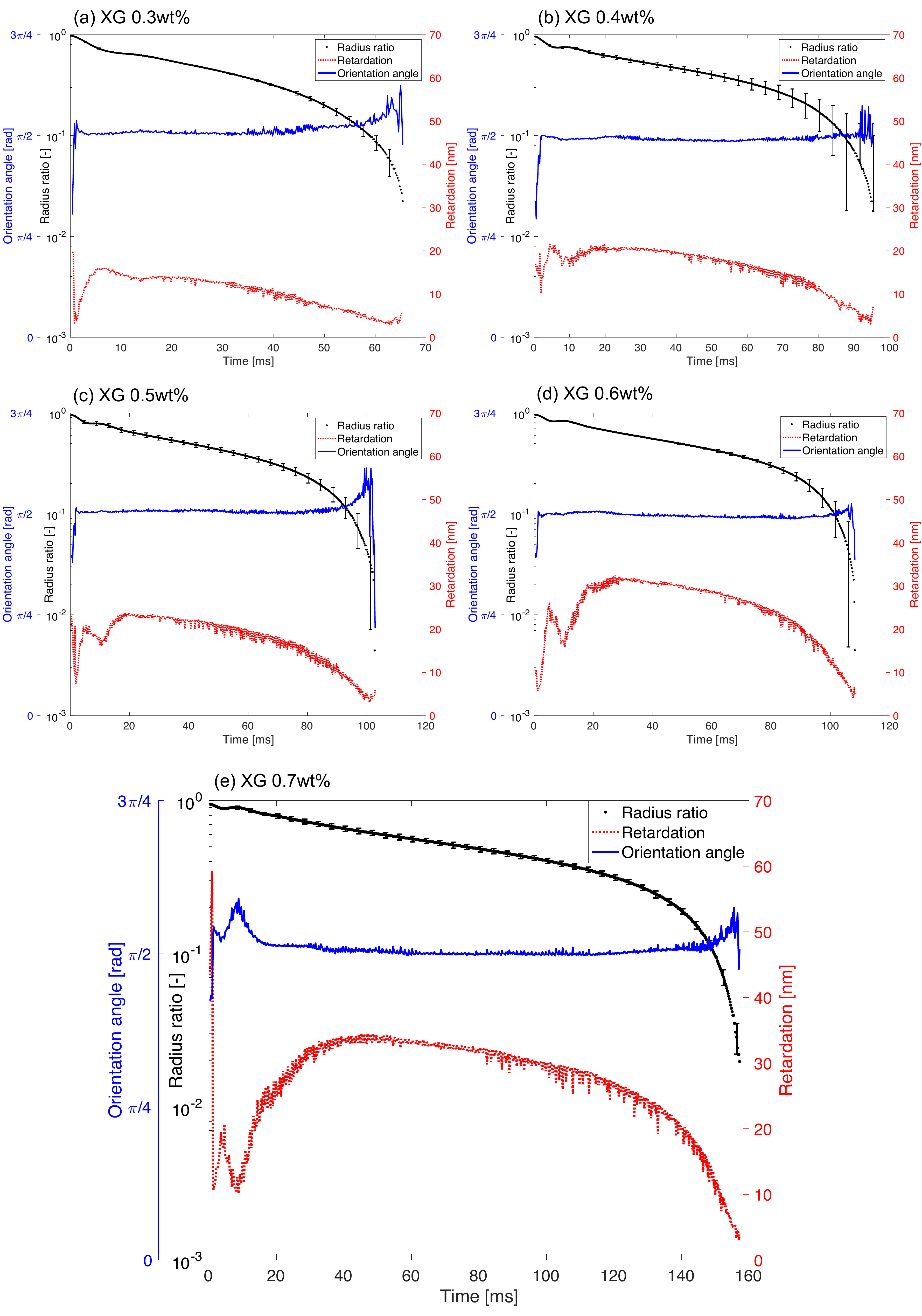}
\caption{\label{XG} 
Temporal evolution of retardation $\delta$ and orientation angle $\varphi$ with radius ratio $R_\mathrm{mid}/R_{0}$ for XG solutions with concentrations of (a) 0.3~wt.\,\%, (b) 0.4~wt.\,\%, (c) 0.5~wt.\,\%, (d) 0.6~wt.\,\%, and (e) 0.7~wt.\,\%.}
\end{figure*}
Figure~\ref{XG} shows the results for radius ratio, retardation, and orientation angle over time for five different concentrations of XG. Within the EC regime, it is known that the liquid filaments of common viscoelastic fluids are cylindrical, while the temporal images of the XG solutions (Fig.~\ref{retardation field XG}) clearly show that the filament decays with a curvature unlike that in the case of PEO solutions.
Thus, it is difficult to determine the presence of the EC regime.
However, since the radius ratio of the XG solutions obeys the scaling law of the EC regime [Eq.~( \ref{radius in ECregime})], we assume the existence of the EC regime here and estimate the durations of the EC regime for XG solutions of different concentrations as follows: 15--30~ms for 0.3~wt.\,\%, 30--60~ms for 0.4~wt.\,\%, 30--60~ms for 0.5~wt.\,\%, 30--60~ms for 0.6~wt.\,\%, and 50--100~ms for 0.7~wt.\,\%.

The results for the orientation angle in Fig.~\ref{XG} confirm that the orientation angle $\varphi$ remains constant at $\pi/2$ except during the early stage (part of the IC regime) and the final stage (part of the VC regime).
Since XG is a semirigid polymer while PEO is a flexible polymer, we initially expected that it would be more difficult to orient the XG than the PEO.
However, in our experiment, both types of polymers were oriented along the extensional direction, and thus it is confirmed that CaBER-DoS provides sufficient extensional stress to orient the polymer chains.
For the retardation of XG solutions (Fig.~\ref{XG}), the curves are clearly different from those of the mixed solutions of PEO and CNCs (Fig.~\ref{PEO}).
The curves for the XG solutions increase in the early stage and then decrease gradually from the EC to the VC regime.

\begin{figure*}[t!]
\includegraphics[width=0.8\textwidth]{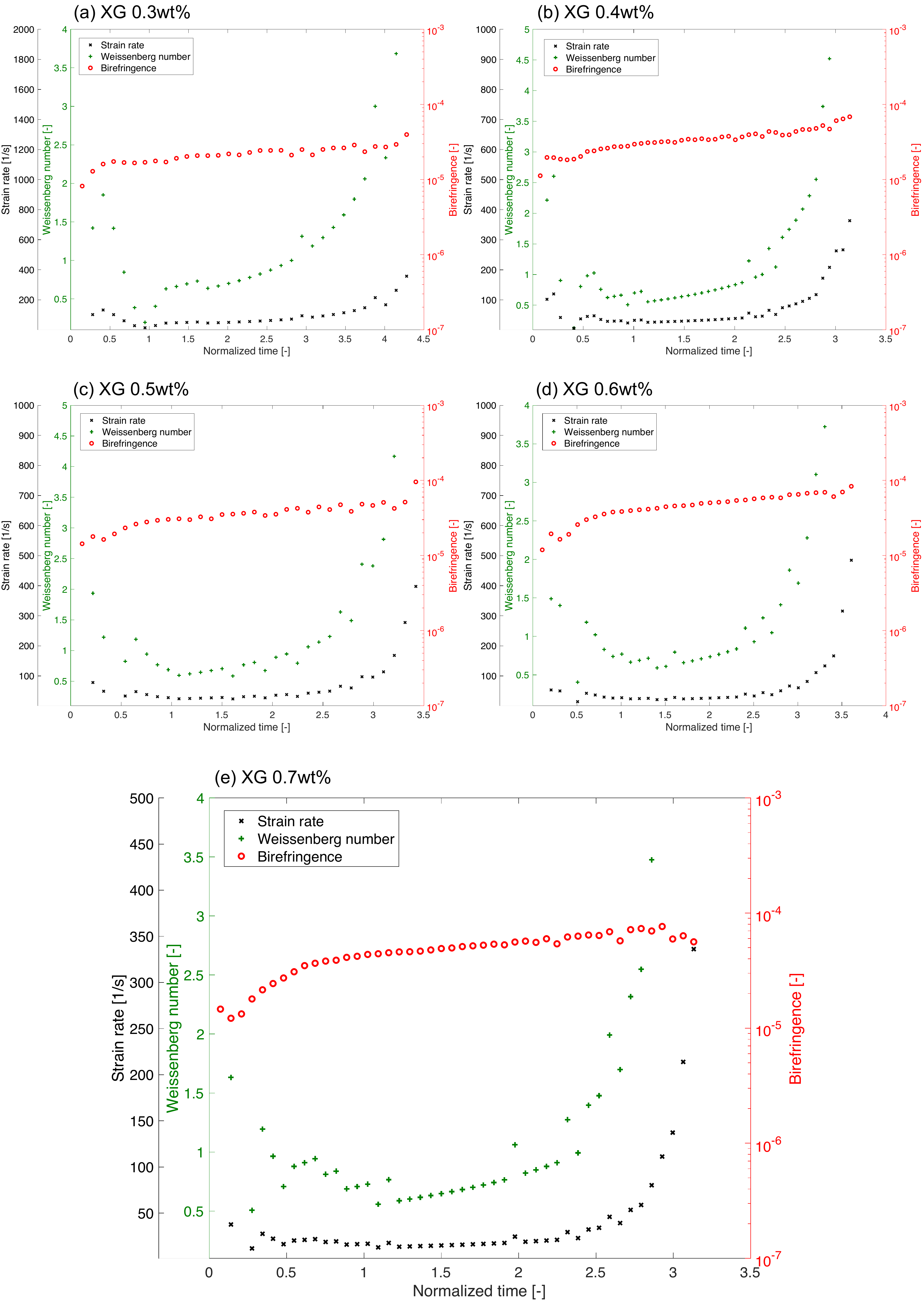}
\caption{\label{birefringence XG} 
Temporal evolution of birefringence $\delta_{n}$, Weissenberg number $Wi$, and strain rate $\dot{\epsilon}_\mathrm{mid}$ for XG solutions with concentrations of (a) 0.3~wt.\,\%, (b) 0.4~wt.\,\%, (c) 0.5~wt.\,\%, (d) 0.6~wt.\,\%, and (e) 0.7~wt.\,\%.}
\end{figure*}
Figure~\ref{birefringence XG} show plots of birefringence $\delta_{n}$ vs normalized time $\hat{t}$.
It is confirmed that the slope of the birefringence for the XG solutions is slightly different at the start of the EC regime, showing a slight increase.
However, we cannot determine whether the birefringence can be an indicator for determining the presence of the EC regime, because it is difficult to extract any significant trend at the transition from the IC to the EC regime (this is evident from the data on $Wi$).
In other words, the birefringence results reflect the fact that the presence of the EC regime cannot be determined from the temporal evolution image of the XG solution, as mentioned above.
We believe this this to be because XG is a naturally extracted polymer, with a nonuniform molecular weight.

\section*{References}
\bibliography{Paper97294_references_EDITED}

\end{document}